\documentclass[twocolumn]{revtex4-1}

\usepackage[T1]{fontenc}
\usepackage{color}
\usepackage{amsmath}
\usepackage{graphicx}
\usepackage{fdsymbol}
\usepackage{epstopdf}

\makeatletter
\providecommand{\tabularnewline}{\\}
\makeatother

\begin{document}
\title{Theta-gamma cross-frequency coupling enables covariance between distant brain regions}

\author{Akihiko Akao}
\affiliation{Graduate School of Engineering, The University of Tokyo, 7-3-1 Hongo, Bunkyo-Ku, Tokyo 113-0033, Japan}

\author{Sho Shirasaka}
\affiliation{Graduate School of Information Science and Technology, 1-5 Yamadaoka, Suita, Osaka 565-0871, Japan}

\author{Yasuhiko Jimbo}
\affiliation{Graduate School of Engineering, The University of Tokyo, 7-3-1 Hongo, Bunkyo-ku, Tokyo 113-0033, Japan}

\author{Bard Ermentrout}
\affiliation{Department of Mathematics, University of Pittsburgh, Pittsburgh, Pennsylvania 15260, USA}

\author{Kiyoshi Kotani}
\affiliation{Research Center for Advanced Science and Technology, The University of Tokyo, 4-6-1 Komaba, Meguro-ku, Tokyo 153-8904, Japan}
\affiliation{JST, PRESTO, 4-1-8 Honcho, Kawaguchi-shi, Saitama 332-0012, Japan}

\begin{abstract}

Cross-frequency coupling (CFC), where the amplitude of a fast neuronal oscillation is modulated by a second slower frequency, is thought to play an important role in long-range communication across distant brain regions.
However, neither the mechanism of its generation nor the influence on spiking dynamics is well understood. Here, we investigate the multiscale dynamics of two interacting distant neuronal modules coupled by inter-regional long-range connections. 
Each neuronal module comprises an excitatory and inhibitory population of quadratic integrate-and-fire neurons connected locally with conductance-based synapses. The two modules are then coupled reciprocally with delays that represent the conduction times over the long distance between them. By assuming a Lorentzian distribution to the probability density function of the membrane potential, we are able to apply the Ott-Antonsen ansatz to reduce the corresponding mean field equations of the spiking dynamics to a small set of delay differential equations.
Bifurcation analysis on these mean field equations shows that inter-regional conduction delay is sufficient to produce CFC via a torus bifurcation, as well as a gamma oscillation via a Hopf bifurcation.
Spike correlation and covariance analysis during the CFC revealed that several local clusters in excitatory population exhibit synchronized firing in gamma-band frequencies.
These clusters exhibit locally decorrelated firings between the cluster pairs within the same population because of their different firing frequencies.
In contrast, the same clusters exhibit long-range gamma-band cross-covariance between the corresponding cluster in the distant populations that have similar firing frequency.
The interactions of the different gamma frequencies in each module produce a beat leading to population-level CFC. 
In order to investigate the impact of CFC on neuronal spike timings, we analyzed spike counts in relation to the phases of the macroscopic fast and slow oscillations of the mean membrane potential. We found population spike counts vary with respect to macroscopic phases. Such firing phase preference accompanies a phase window with high spike count and low Fano factor, which is suitable for a population rate code. In addition, we analyzed the firing phase preference of the local clusters. We found these clusters also exhibit firing phase preference that differs between the clusters, similar to experimental findings.
Our work suggests that the inter-regional conduction delay plays a significant role in the emergence of CFC and the underlying spiking dynamics may support long-range communication and neural coding.

\end{abstract}

\maketitle

\section{Introduction}

Rhythms and neuronal oscillations are ubiquitous in the central nervous system (CNS) where they are believed to play a role in many functions ranging from cognition to motor function \cite{buzsaki_book,wang_rev}. These macroscopic rhythms (seen, for example, in electroencephalogram recordings or in extracellular recordings of the local field potential)  emerge from the interactions of populations of excitatory and inhibitory neurons that comprise the cortex and other CNS regions. Thus, there have been many papers written on the biophysical mechanisms that underlie these ubiquitous rhythms \cite{buz-wang}.   

Cross-frequency coupling(CFC), which is a phenomena where a low frequency oscillation modulates the amplitude of a high frequency oscillation, has received a good deal of recent attention experimentally \citep{Jensen_CFC_rev,canolty_CFC_rev,Juhan_CFC_rev,mizureki_CFC_tgcouple} and theoretically\citep{CFC_Gutkin_rev,CFC_sase,CFC_Hyafil_speech,CFC_plone_1,CFC_plone_2}. It has been suggested that high frequency neuronal oscillations may contribute to local computations, while low frequency oscillations are employed in long-range communication across different brain regions\citep{brainweb,CFC_Gutkin_rev}. Therefore, CFC is assumed to play some role in integrating local computations distributed across the distant brain regions and thus facilitating higher cognitive function. This assumption, in which CFC contributes to long-range integration, is supported by some recent experimental studies\citep{IR_CFC_1,IR_CFC_2,IR_CFC_3,IR_CFC_4}. However, the dynamical mechanism for the emergence of CFC during long-range neuronal interactions and the dynamical characteristics of such CFC are still unclear.

To elucidate the mechanisms for  macroscopic neuronal dynamics, large scale mathematical models of neuronal populations have been frequently utilized \citep{Breakspear}. 
Conventional firing rate models are popular to describe macroscopic rhythms\citep{WC_wcmodel,Bard_mathematical_foundataion}. However, they assume that underlying spiking dynamics is asynchronous, which is not suitable for high frequency oscillations \citep{Bard_mathematical_foundataion,Devalle_spike_sychrony,temporal_deccoration}. Recently, the derivation of mean field equations for a population of theta neurons (or the equivalent quadratic integrate-and-fire neurons) via the Ott-Antonsen ansatz\citep{Ott_Antonsen_OA} has been applied \citep{montbrio_OA_QIF}. This approach allows one to capture the synchronized spiking dynamics as well as high frequency oscillations\citep{Devalle_spike_sychrony}. In previous work, pulse-coupling\citep{Luke_complete_pulse}, gap junctions\citep{Laing_Gap_junction}, locally connected inhibitory population\citep{Pazo_QPS2chaos_delay}, a large population \citep{Devalle_a_large}, excitatory and inhibitory (E-I) populations\citep{Dumont_OA_PRC} and time-varying modulation\citep{Paulso_time_varying} have all been considered using this approach, and behavior including, in particular, the emergence of macroscopic oscillations and chaos has been investigated. However, this prior research focused on the local dynamics without long-range interactions and CFC was also not described in these studies. In addition, these works were performed with relatively simple models, where the biological factors which are known to affect the fast macroscopic oscillation such as synaptic conductance dynamics\citep{brunel_what_determines} and synaptic reversal potential\citep{reversal_V_affects_gamma,revV_affect_2} have not been fully incorporated.

In our previous work, we introduced a modified quadratic integrate-and-fire neuron and analyzed the emergence of (fast) gamma oscillations and analyzed their macroscopic phase response curves\citep{kotani_ing,akao_pre}. However, because our focus was on the local emergence of high frequency oscillations, the dynamics during long-range interactions between distant brain regions was not analyzed.

Thus, here, we analyze the dynamics of two delay-coupled E-I modules composed of modified quadratic integrate-and-fire (QIF) neurons. We adopt the Ott-Antonsen ansatz for each population and extract their mean field dynamics. We introduce the long range interaction through a synaptic conduction delay between the two distant E-I modules and derive the mean field dynamics of the whole system as a set of delay differential equations (DDEs). We perform bifurcation analysis on the mean field DDEs and investigate how the macroscopic oscillations are facilitated from these long-range interactions.

We first introduce the model and the resulting Ott-Antonsen reduction. We show that a conduction delay between the regions is sufficient to produce CFC and then show that the CFC arises in the mean field model as a torus bifurcation near the intersection of two distinct Hopf bifurcation branches. We analyze the spiking model in the parameter regions where there is CFC and show that it induces long-range multi-cluster gamma-band cross-covariance. In addition, we find spike timings of individual neurons have a preference for specific phases of both the theta and the gamma oscillations, thus generating a suitable temporal window for a population rate code. Our results suggest that CFC is one of the consequences of inter-regional coupling and that the underlying spiking dynamics may support long-range information integration and neural coding.

\section{Methods}

\subsection{Dynamics of the spiking neuronal population}

We consider two pairs of E-I modules as shown in Fig. \ref{fig:Delay-coupled-populations-of}(a), namely module $1$ and $2$. In each module, we consider E and I populations of the modified QIF neurons\citep{kotani_ing,akao_pre}, namely $E_{1}$, $I_{1}$, $E_{2}$ and $I_{2}$.

The dynamics of the membrane potential $V_{i}^{X}(t)$ , of the $i$-th neuron in population $X$ is written as

\begin{align}
C\frac{dV_{i}^{X}}{dt}= & g_{LX}\frac{(V_{i}^{X}(t)-V_{R})(V_{i}^{X}(t)-V_{T})}{V_{T}-V_{R}}\nonumber \\
 & -\sum_{Y}g_{X}^{Y}(t)(V_{i}^{X}(t)-V_{syn}^{Y})+I_{i}^{X}.\label{eq:V_N}
\end{align}

Parameters are listed in Tab. \ref{tab:Parameters-of-qif}. $V_{i}^{X}(t)$ follows a resetting rule as $if\ V_{i}^{X}(t)\geq V_{peak},\ then\ V_{i}^{X}(t)\leftarrow V_{reset}$. Such resetting of $V_{i}^{X}(t)$ represents the action potential or firing of neuron $i$.

\begin{table*}
\caption{Parameters used Eq.\ref{eq:V_N}. \label{tab:Parameters-of-qif}}

\centering{}%
\begin{tabular}{|c|c|c|}
\hline 
Symbol  & Property  & Default value (Unit)\tabularnewline
\hline 
\hline 
$X$  & post-synaptic population  & $X=\left\{ I_{1},I_{2},E_{1},E_{2}\right\} $\tabularnewline
\hline 
$Y$  & pre-synaptic population  & $Y=\left\{ I_{1},I_{2},E_{1},E_{2}\right\} $\tabularnewline
\hline 
\hline 
$N_{X}$  & number of neurons  & %
\begin{tabular}{c}
$N_{I_{1}}=N_{I_{2}}=100$ \tabularnewline
$N_{E_{1}}=N_{E_{2}}=400$\tabularnewline
\end{tabular}\tabularnewline
\hline 
$C$  & membrane capacitance  & $C=1\mathrm{(\mu F/cm^{2})}$ \tabularnewline
\hline 
$i$  & index of neuron  & $1\leq i\leq N_{X}$\tabularnewline
\hline 
$g_{LX}$  & leak conductance  & %
\begin{tabular}{c}
$g_{LI_{1}}=g_{LI_{2}}=0.1\mathrm{(mS/cm^{2})}$\tabularnewline
$g_{LE_{1}}=g_{LE_{2}}=0.08\mathrm{(mS/cm^{2})}$\tabularnewline
\end{tabular}\tabularnewline
\hline 
$V_{R}$  & resting potential  & $V_{R}=-62$(mV)\tabularnewline
\hline 
$V_{T}$  & threshold potential  & $V_{T}=-55$(mV)\tabularnewline
\hline 
$I_{i}^{X}$  & quenched current  & %
\begin{tabular}{c}
Distributed among population $X$\tabularnewline
following the Lorentz distribution\tabularnewline
$f_{X}(I_{X})=\frac{1}{\pi}\frac{\Delta_{X}}{\left(I-\bar{I}_{X}\right)^{2}+\Delta_{X}^{2}}$\tabularnewline
\end{tabular}\tabularnewline
\hline 
$\bar{I}_{X}$  & median of $I_{i}^{X}$  & $\bar{I}_{E_{1}}=\bar{I}_{E_{2}}=\bar{I}_{I_{1}}=\bar{I}_{I_{2}}=1\mathrm{(A/cm^{2})}$\tabularnewline
\hline 
$\Delta_{I_{X}}$  & HWHM of $I_{i}^{X}$  & $\Delta_{E_{1}}=\Delta_{E_{2}}=\Delta_{I_{1}}=\Delta_{I_{2}}=0.1\mathrm{(A/cm^{2})}$\tabularnewline
\hline 
$V_{syn}^{Y}$  & synaptic reversal potential  & %
\begin{tabular}{c}
$V_{syn}^{I_{1}}=V_{syn}^{I_{2}}=-70$ (mV)\tabularnewline
$V_{syn}^{E_{1}}=V_{syn}^{E_{2}}=0$ (mV)\tabularnewline
\end{tabular}\tabularnewline
\hline 
\end{tabular}
\end{table*}

We assume that the synaptic connections from population $Y$ to $X$ occur with probability $p_{X}^{Y}$. The dynamics of $g_{X}^{Y}(t)$ , synaptic conductance from $Y$ to $X$, is written as 
\begin{equation}
\frac{dg_{X}^{Y}}{dt}=-\frac{1}{\tau_{d}^{Y}}g_{X}^{Y}(t)+\bar{g}_{X}^{Y}\cdot p_{X}^{Y}\cdot\sum_{k=1}^{N_{Y}}\delta\left( t-(  t^{Y_{(k)}}  ) +\tau_{X}^{Y})\right).\label{eq:syn-homo-N-1}
\end{equation}

Parameters are listed in Tab. \ref{tab:Parameters-of-syn} and Tab. \ref{tab:Parameters-of-g_peak}. The conduction delay $\tau_{X}^{Y}$ is $\tau_{delay}$ for the cases where coupling is inter-regional. $\tau_{X}^{Y}$ is zero for the other cases where coupling is local. It has been experimentally shown that the conduction delay between distant brain regions can be up to $40$ (ms) \citep{innocenti_delay_length}. Therefore, we set $\tau_{delay}=31$ (ms) as the nominal default value. For inter-module connections, the synaptic strength $p_{X}^{Y}$ is set as $p_{delay}$. For local connections , the synaptic strength $p_{X}^{Y}$ is set as $p_{local}$ except $p_{I_{2}}^{I_{2}}$ which is the recurrent inhibition in the $I_{2}$ population. $p_{I_{2}}^{I_{2}}$ is set as $p_{local}\times0.9$ in order to introduce a slight difference between module 1 and module 2 (to break any symmetry). We set the area of a neuron as $2.9\times10^{-4}\mathrm{(cm^{2})}$ to match the physiological plausible value\citep{aon_exp,aon}. Peak conductances $\bar{g}_{Y}^{X}$ are also set as to match physiological plausible values \citep{brunel_what_determines,g_peak_1,g_peak_2}.


\begin{table*}
\caption{Parameters used Eq.\ref{eq:syn-homo-N-1}. \label{tab:Parameters-of-syn}}

\centering{}%
\begin{tabular}{|c|c|c|}
\hline 
Symbol  & Property  & Default value (Unit)\tabularnewline
\hline 
\hline 
$\tau_{d}^{Y}$  & decay time constant  & %
\begin{tabular}{c}
$\tau_{d}^{I_{1}}=\tau_{d}^{I_{2}}=5$ (ms)\tabularnewline
$\tau_{d}^{E_{1}}=\tau_{d}^{E_{2}}=2$ (ms)\tabularnewline
\end{tabular}\tabularnewline
\hline 
$\bar{g}_{X}^{Y}$  & peak conductance  & Listed in Tab. \ref{tab:Parameters-of-g_peak}\tabularnewline
\hline 
$k^{Y}$  & index of pre-synaptic neuron  & $1\leq k\leq N_{Y}$\tabularnewline
\hline 
$ t^{Y_{(k)}}$  & spike time of $k$-th neuron in $Y$  & $\mathrm{(ms)}$\tabularnewline
\hline 
$\tau_{X}^{Y}$  & conduction delay from $Y$ to $X$  & %
\begin{tabular}{c}
$0\mathrm{(ms)}$ : coupling within a module\tabularnewline
$\tau_{delay}\mathrm{(ms)}$ : coupling between modules \tabularnewline
\end{tabular}\tabularnewline
\hline 
$\tau_{delay}$  & conduction delay between modules  & $\tau_{delay}=31$ (ms)\tabularnewline
\hline 
$p_{X}^{Y}$  & synaptic strength from $Y$ to $X$  & %
\begin{tabular}{c}
$p_{local}$: coupling within a module (except $p_{I_{2}}^{I_{2}}$)\tabularnewline
$p_{delay}$: coupling between modules\tabularnewline
$p_{I_{2}}^{I_{2}}$: coupling from $I_{2}$ to $I_{2}$\tabularnewline
\end{tabular}\tabularnewline
\hline 
$p_{local}$  & synaptic strength within a module (except $p_{I_{2}}^{I_{2}}$)  & $p_{local}=0.15$\tabularnewline
\hline 
$p_{delay}$  & synaptic strength between modules  & $p_{delay}=0.125$\tabularnewline
\hline 
$p_{I_{2}}^{I_{2}}$  & recurrent synaptic strength in population $I_{2}$  & $p_{I_{2}}^{I_{2}}=0.135$\tabularnewline
\hline 
\end{tabular}
\end{table*}

\begin{table*}
\caption{Peak conductance $\bar{g}_{Y}^{X}$. \label{tab:Parameters-of-g_peak}}

\centering{}%
\begin{tabular}{|c|c|c|}
\hline 
Symbol  & Property  & Default value (Unit)\tabularnewline
\hline 
\hline 
$\bar{g}_{E_{1}}^{E_{1}}=\bar{g}_{E_{2}}^{E_{2}}=\bar{g}_{E_{1}}^{E_{2}}=\bar{g}_{E_{2}}^{E_{1}}$  & AMPA on pyramidal cell  & 4.069$\times10^{-3}$ $\mathrm{(mS/cm^{2})}$\tabularnewline
\hline 
$\bar{g}_{E_{1}}^{I_{1}}=\bar{g}_{E_{2}}^{I_{2}}=\bar{g}_{E_{1}}^{I_{2}}=\bar{g}_{E_{2}}^{I_{1}}$  & AMPA on interneuron  & 3.276$\times10^{-3}$ $\mathrm{(mS/cm^{2})}$\tabularnewline
\hline 
$\bar{g}_{I_{1}}^{E_{1}}=\bar{g}_{I_{2}}^{E_{2}}$  & GABA on pyramidal cell  & 2.672$\times10^{-2}$ $\mathrm{(mS/cm^{2})}$\tabularnewline
\hline 
$\bar{g}_{I_{1}}^{I_{1}}=\bar{g}_{I_{2}}^{I_{2}}$  & GABA on interneuron  & 2.138$\times10^{-2}$ $\mathrm{(mS/cm^{2})}$\tabularnewline
\hline 
\end{tabular}
\end{table*}

The description of the whole system (spiking neuronal population and synaptic dynamics) is obtained by a set of 1000 of Eq. \ref{eq:V_N} (400 E and 100 I cells for each area) and 12 of Eq. \ref{eq:syn-homo-N-1}.

For the numerical simulation of the spiking population, we transformed Eq. \ref{eq:V_N} to the form of the theta neuron to avoid numerical delicacies near the spiking threshold. We introduce the following transformation:
\begin{equation}
V_{i}^{X}=\frac{V_{R}+V_{T}}{2}+\frac{V_{T}-V_{R}}{2}\tan\frac{\theta_{i}^{X}}{2}.\label{eq:theta-v-map_individual}
\end{equation}
We then take the limit as $V_{peak}$ =$-V_{reset}$ $=+\infty$, which naturally allows us to capture the neuronal spike as well as the refractory period that is evoked by the spike. Then, Eq.\ref{eq:V_N} can be transformed into 
\begin{align}
C\frac{d}{dt}\theta_{i}^{X}(t) & =-g_{LX}\cos\theta_{i}^{X}(t)+h(1+\cos\theta_{i}^{X}(t))I_{i}^{X}\nonumber \\
 & +\sum_{Y}g_{Y}^{X}(t)\left\{ q^{Y}(1+\cos\theta_{i}^{X}(t))-\sin\theta_{i}^{X}(t)\right\} ,\label{eq:theta}
\end{align}

where $h=2/\left(V_{T}-V_{R}\right)$ and $q^{Y}=\left(2V_{syn}^{Y}-V_{R}-V_{T}\right)/\left(V_{T}-V_{R}\right)$. Note that $V=\pm\infty$ in Eq. \ref{eq:V_N} corresponds to $\theta=\pm\pi$ in Eq.\ref{eq:theta}.

We used a set of 1000 of Eq. \ref{eq:theta} and 12 of Eq. \ref{eq:syn-homo-N-1} to numerically simulate the dynamics of the spiking population.

\subsection{Dynamics of mean field equations}

We adopt the Ott-Antonsen ansatz\citep{Ott_Antonsen_OA}, which allows us to obtain a mean field description of the spiking model in the limit as $N\to\infty$. Following the previous work\citep{montbrio_OA_QIF}, we get  mean field equations for Eq. \ref{eq:V_N} and Eq.\ref{eq:syn-homo-N-1} as:

\begin{align}
\frac{d}{dt}r_{X} & =2a_{X}r_{X}(t)v_{X}(t)+b_{X}(t)r_{X}(t)+\frac{a_{X}}{\pi}\Delta_{I_{X}},\label{eq:r-dynamics}
\end{align}

\begin{align}
\frac{d}{dt}v_{X} & =a_{X}v_{X}(t)^{2}-\frac{\pi^{2}}{a_{X}}r_{X}(t)^{2}+b_{X}(t)v_{X}(t)+c_{X}(t)+\bar{I_{X}},\label{eq:v-dynamics}
\end{align}
\begin{equation}
\frac{dg_{X}^{Y}}{dt}=-\frac{1}{\tau_{d}^{Y}}g_{X}^{Y}(t)+\bar{g}_{X}^{Y}\cdot p_{X}^{Y}\cdot N_{Y}\cdot r_{Y}(t-\tau_{X}^{Y}),\label{eq:syn-homo-N-2}
\end{equation}

where $a_{X}=g_{LX}/\left(V_{T}-V_{R}\right)$, $b_{X}\left(t\right)=-g_{LX}\left(V_{T}+V_{R}\right)/\left(V_{T}-V_{R}\right)-\sum_{Y}g_{X}^{Y}(t)$ and $c_{X}\left(t\right)=-g_{LX}V_{T}V_{R}/\left(V_{T}-V_{R}\right)+\sum_{Y}g_{X}^{Y}(t)V_{syn}^{Y}$. The details of the derivation are in Appendix \ref{subsec:Derivation-of-mean}. We will use this mean field equation to investigate the emergence of macroscopic oscillations by bifurcation analysis.

\subsection{Bifurcation analysis of the mean field equations}

We use DDE-BIFTOOL to analyze the existence and stability of the solutions to Eqs. \ref{eq:r-dynamics}, \ref{eq:v-dynamics} and \ref{eq:syn-homo-N-2}. The numerical methods used in DDE-BIFTOOL are detailed in \citep{ddebif_1,ddebif_2,ddebif_3,ddebif_4,ddebif_5,ddebif_6}.

\section{Results}

\subsection{Conduction delay induces cross-frequency coupling}

\begin{figure}[htp]

\includegraphics[width=1\linewidth]{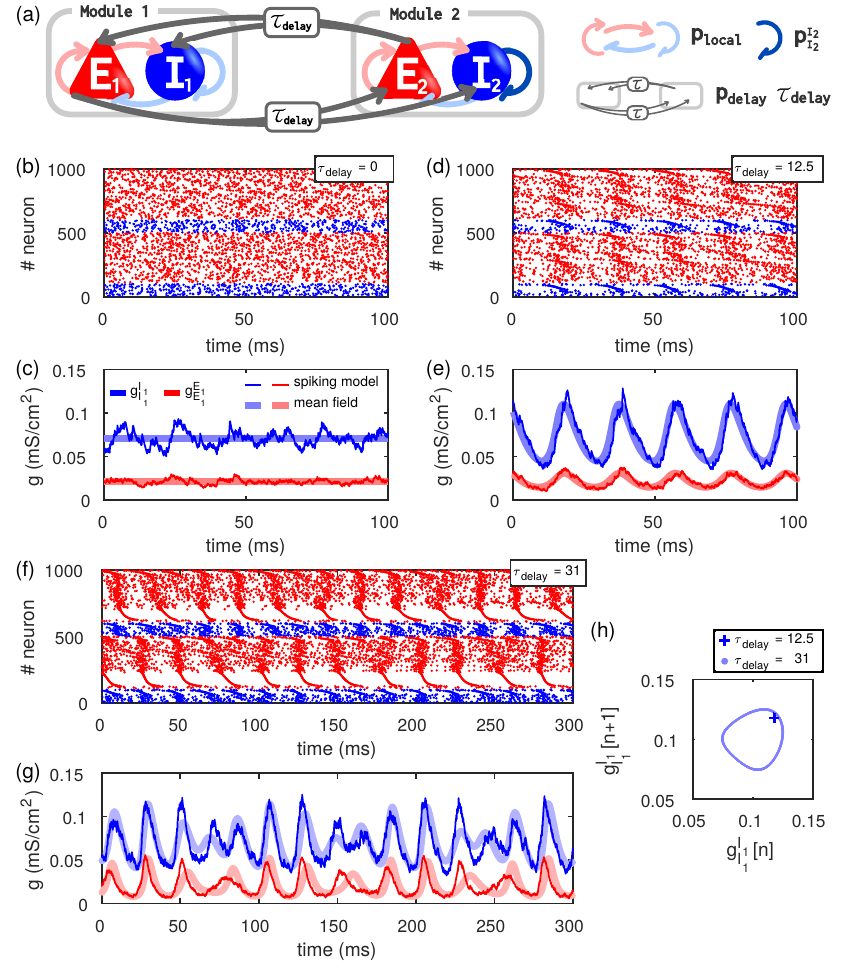}

\caption{Population dynamics of delay-coupled modified theta neurons. (a) Schematics. The coupling between the two modules is delayed because of long-range conduction time. (b-g) Result of simulations. $p_{local}=0.15$, $p_{delay}=0.125$ and $p_{I_{2}}^{I_{2}}=0.135$. (b,c) $\tau_{delay}=0$ (ms). (d,e) $\tau_{delay}=12.5$ (ms). (f,g) $\tau_{delay}=31$ (ms). (b,d,f) Rastergram. (c, e, g) Time series of $g_{I_{1}}^{I_{1}}$ and $g_{E_{1}}^{E_{1}}$. The dynamics of spiking population and mean field equations are in good agreement. (h) Lorenz map of the peak of $g_{I_{1}}^{I_{1}}(t)$.\textcolor{red}{{} \label{fig:Delay-coupled-populations-of}}}
\end{figure}

We performed three sets of numerical simulations where $\tau_{delay}$ has values of $0\mathrm{(ms)}$, $12.5\mathrm{(ms)}$ and $31\mathrm{(ms)}$ to investigate the effect of the inter-module conduction delay on the dynamics. Parameters except $\tau_{delay}$ are fixed at the default values shown in Tabs. \ref{tab:Parameters-of-qif} - \ref{tab:Parameters-of-g_peak}. Simulation was performed in both the spiking population (Eqs. \ref{eq:syn-homo-N-1} and \ref{eq:theta}) and the mean-field equations (Eqs. \ref{eq:r-dynamics}, \ref{eq:v-dynamics} and \ref{eq:syn-homo-N-2}). The two corresponding results were compared to confirm the validity of the mean field equations in the presence of the delay.
(We note that 
the validity of the Ott-Antonsen approach has 
not been formally proven in the case where there are delays, although it has been empirically shown to work for delayed cases \citep{LeeDelay,Pazo_QPS2chaos_delay,Devalle_a_large}.) When $\tau_{delay}=0\mathrm{(ms)}$, no macroscopic rhythm is observed. The spikes are asynchronous {[}Fig. \ref{fig:Delay-coupled-populations-of}(b){]} and the synaptic conductances are constant values, although we can see fluctuations due to the finite size effect in the spiking population {[}Fig. \ref{fig:Delay-coupled-populations-of}(c){]}.
When $\tau_{delay}=12.5\mathrm{(ms)}$, a gamma oscillation is observed. The spikes are partially synchronous {[}Fig. \ref{fig:Delay-coupled-populations-of}(d){]} and the synaptic conductances exhibits oscillations at around 50Hz {[}Fig. \ref{fig:Delay-coupled-populations-of}(e){]}, which can be regarded as a gamma oscillation.
When $\tau_{delay}=31\mathrm{(ms)}$, the spikes are partially synchronous {[}Fig. \ref{fig:Delay-coupled-populations-of}(f){]}. The synaptic conductances are oscillating in the gamma range {[}Fig. \ref{fig:Delay-coupled-populations-of}(g){]}. Moreover, these dynamics can be regarded as a form of  CFC because the amplitude of the gamma oscillation is modulated by a slow rhythm. Also, this CFC can be called  ``theta-gamma coupling\textquotedblright{} because the fast oscillation is about 50Hz which is in the gamma range and the slow modulation is about 10 Hz which is in the theta range.
These simulations show that the inter-module delay $\tau_{delay}$ drastically affects the dynamics, especially on the emergence of the rhythm. We also note that both the gamma oscillation and CFC emerge just by increasing the time-delay in our model.

We found that the gamma oscillation was a limit cycle and the CFC was a torus. This is indicated in a Lorenz plot of $g_{I_{1}}^{I_{1}}$, which is a plot of $(g_{I_{1}}^{I_{1}}[n],g_{I_{1}}^{I_{1}}[n+1])$ in x-y plane where $g_{I_{1}}^{I_{1}}[n]$ is the $n$-th local maximum value of $g_{I_{1}}^{I_{1}}$ {[}Fig. \ref{fig:Delay-coupled-populations-of}(h){]} \citep{strogatz}. Here, $g_{I_{1}}^{I_{1}}$ is simulated with the mean field equations (Eqs. \ref{eq:r-dynamics}, \ref{eq:v-dynamics} and \ref{eq:syn-homo-N-2}). When $\tau_{delay}=12.5\mathrm{(ms)}$, the Lorenz plot is a point, which indicates the dynamics is a limit cycle. When $\tau_{delay}=31\mathrm{(ms)}$, the Lorenz plot is a circle which indicates the dynamics is a torus\citep{Lorenz}.The bifurcation analysis of these attractors will be investigated in the next session.
We also show the corresponding mean field (Eqs. \ref{eq:r-dynamics}, \ref{eq:v-dynamics} and \ref{eq:syn-homo-N-2}) behavior along with that of the spiking model in {[}Figs. \ref{fig:Delay-coupled-populations-of}(c), (e) and (g){]}. The dynamics of the mean field equations are generally consistent with the corresponding spiking population (Eqs. \ref{eq:theta} and \ref{eq:syn-homo-N-1}) despite the fluctuations due to the finite size of the spiking model.

\subsection{Cross-frequency coupling by delay-induced torus bifurcation}

\begin{figure}[htp]

\includegraphics[width=1\linewidth]{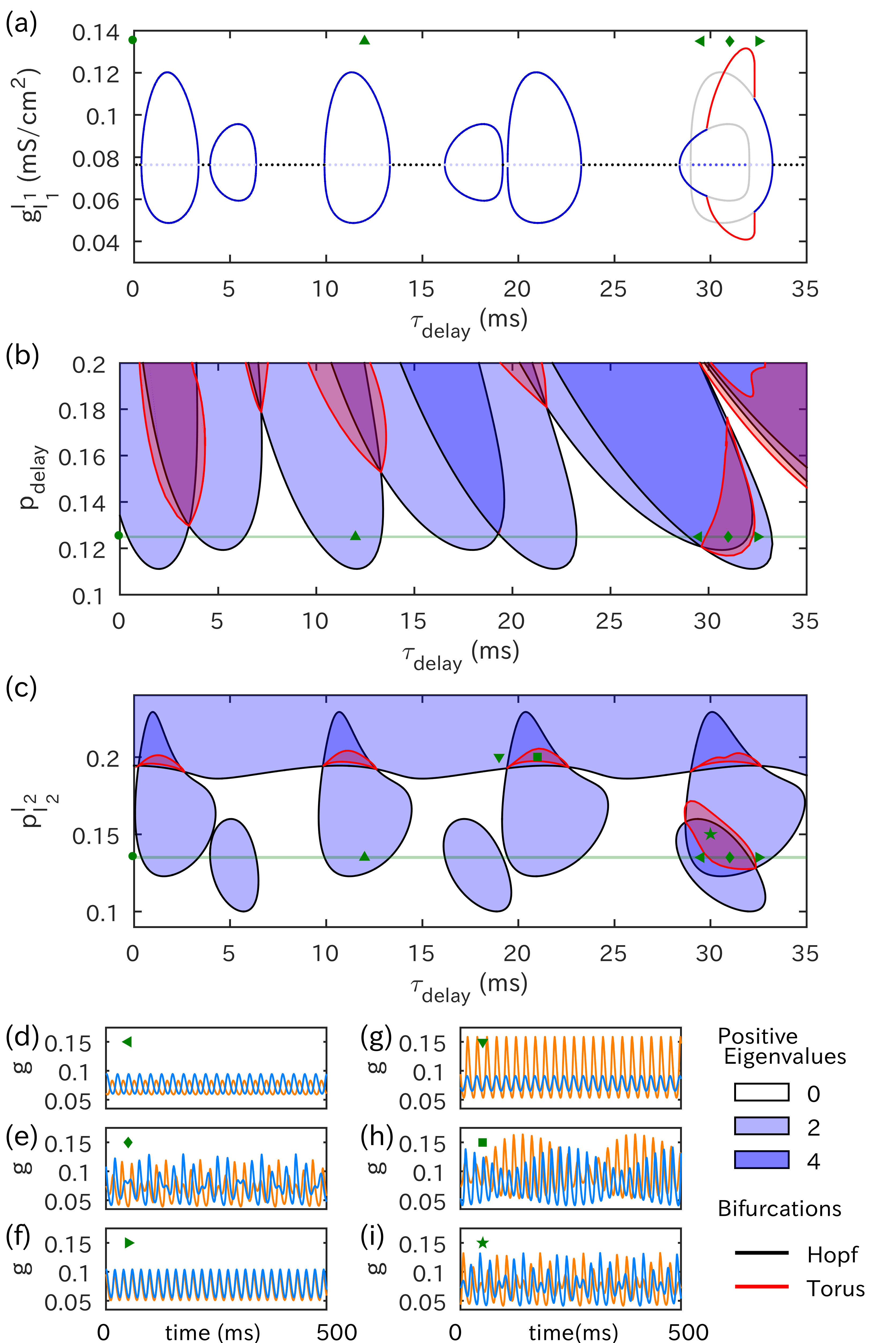}
\caption{Bifurcation diagrams (a-c) and simulations (d-i). (a) One parameter bifurcation diagram with $\tau_{delay}$ and $g_{I_{1}}^{I_{1}}$. Black dotted line indicates stable steady state, light blue dotted line is one pair of eigenvalues with positive real parts and dark blue dotted line is four positive real part eigenvalues.  Solid blue line indicates stable limit cycle orbit and solid gray line indicates unstable limit cycle orbit (maximum and minimum values). Red line indicates the torus orbit (maximum and minimum values, computed by integration of the system). In the following bifurcation diagrams, parameters used in Fig. \ref{fig:Delay-coupled-populations-of} (b,c), Fig. \ref{fig:Delay-coupled-populations-of} (d,e) and Fig. \ref{fig:Delay-coupled-populations-of} (f,g) are marked as $\bigcirc$ , $\triangle$ and $\diamondsuit$, respectively. (b) Two parameter diagram with $\tau_{delay}$ and $p_{delay}$. The Hopf bifurcation and torus bifurcation are plotted. Also, the number of positive eigenvalues are indicated by shaded blue area. The region of a stable torus solution is indicated by red shaded area. (c) Two parameter bifurcation diagram with $\tau_{delay}$ and $p_{I_{2}}^{I_{2}}$. Similarly, the bifurcation boundaries and the number of positive eigenvalues are plotted. (d-i) Result of numerical simulation with parameter sets marked in Fig. \ref{fig:Bifurcation-diagrams} (a-c). Blue: $g_{E_{1}}^{E_{1}}$. Orange: $g_{E_{2}}^{E_{2}}$. (d) $\triangleleft$: $\tau_{delay}= 29$ (ms) (e) $\diamondsuit$: $\tau_{delay}= 31$ (ms) (f) $\triangleright$: $\tau_{delay}= 33$ (ms) (g) $\bigtriangledown$: $\tau_{delay}=19$ (ms), $p_{I_2}^{I_2}=0.2$ (h) $\Box$: $\tau_{delay}=22$ (ms), $p_{I_2}^{I_2}=0.2$ (i) $\medwhitestar$: $\tau_{delay}=30$ (ms), $p_{I_2}^{I_2}=0.15$. \textcolor{red}{\label{fig:Bifurcation-diagrams}}}
\end{figure}

We performed bifurcation analysis on the mean field equations to reveal the mechanism for the qualitative difference of the dynamics that we found in Fig. \ref{fig:Delay-coupled-populations-of}. We show three bifurcation diagrams in Figs. \ref{fig:Bifurcation-diagrams} (a-c).

Fig. \ref{fig:Bifurcation-diagrams} (a) is a one-parameter bifurcation diagram where the horizontal axis is  $\tau_{delay}$, and the vertical axis is the value of the equilibrium point or the envelope of the oscillation for the dynamical variable $g_{I_{1}}^{I_{1}}$. 
Parameters used in Fig. \ref{fig:Delay-coupled-populations-of} (b,c), Fig. \ref{fig:Delay-coupled-populations-of} (d,e) and Fig. \ref{fig:Delay-coupled-populations-of} (f,g) corresponds to circle, triangle and diamond, respectively.
We can see limit cycles emerge via Hopf bifurcations (HB), and a torus emerges in a parameter region where two distinct limit cycle orbits coexist. For $\tau_{delay}=0$(ms), the black dotted line indicates the equilibrium point is stable and the light blue dotted lines indicate unstable equilibrium. As $\tau_{delay}$ increases, a limit cycle oscillation emerges from a HB point at $\tau_{delay}=0.39$(ms). The limit cycle ends with another HB at $\tau_{delay}=3.4$(ms). For larger $\tau_{delay}$, another HB is induced at $\tau_{delay}=4.0$(ms) and ends with $\tau_{delay}=6.4$(ms). As $\tau_{delay}$ increases, limit cycles with large and small amplitudes appear regularly in turn with different intervals. As the consequence of their regular appearance, around $\tau_{delay}=30$(ms), two limit cycles coexist. In this region, we can see the limit cycles become unstable, and a torus emerges via a torus bifurcation. Unlike the limit cycles and fixed point, which are computed via continuation with DDE-BIFTOOL, the red curve showing the tori is computed by direct forward integration of the mean field equations.

Fig. \ref{fig:Bifurcation-diagrams}(b) and (c) are two-parameter bifurcation diagrams which show the region of equilibrium solutions, limit cycles, and tori in a 2D parameter space. The horizontal axis is for the parameter $\tau_{delay}$. The vertical axis is for $p_{delay}$(b) and $p_{I_{1}}^{I_{1}}$(c).

In Fig. \ref{fig:Bifurcation-diagrams}(b), we found CFC emerges as a torus in the area surrounded by two torus branches emanate from the double-Hopf bifurcation (DHB) point where two HB boundaries overlap. In Fig. \ref{fig:Bifurcation-diagrams}(b), the green line of $p_{delay}=0.135$ corresponds to the parameters shown in Fig. \ref{fig:Bifurcation-diagrams}(a). We can also see the HB region periodically emerges. These HB regions become wider as $p_{delay}$ becomes large, resulting in U-shaped HB boundaries. We can see two distinct HB boundaries overlap at some points, which are the DHB points where the complex eigenvalues of the two sets (total of 4) exist on the imaginary axis. As shown in a previous analysis of this codimension 2 bifurcation, a two-torus branches emanate from a DHB point \citep{buono,double-hopf}. The area surrounded by the torus branch shown above is coincident with the area where the torus occurs in Fig. \ref{fig:Bifurcation-diagrams} (a). From these facts, we have shown that the generation mechanism of CFC is due to the torus bifurcations and it emanates from the DHB point where two HB branches overlap.

In Fig. \ref{fig:Bifurcation-diagrams} (c), gamma oscillations emerge when $p_{I_{2}}^{I_{2}}$ is large enough ($p_{I_{2}}^{I_{2}}>0.2$), regardless of the values of $\tau_{delay}$ . The emergence of gamma oscillations due to strong local inhibition has also been reported in previous studies. When $p_{I_{2}}^{I_{2}}\leq0.2$, stable  gamma oscillations and CFC coexist depending on $p_{I_{2}}^{I_{2}}$ and $\tau_{delay}$. We note CFC emerges in networks even with symmetric coupling between modules (i.e. $p_{I_{2}}^{I_{2}}=p_{local}=0.15$). For example, when $\tau_{delay}=30$ (ms) as indicated by a star, it is inside the torus bifurcation region, and the emergence of the torus is confirmed also from the numerical simulation. In Fig. \ref{fig:Bifurcation-diagrams} (d-i), the numerical simulations of the representative parameters are shown. Since the qualitative features of them agree with the prediction of the bifurcation diagram, this validates the bifurcation analysis of the mean field model.

\subsection{Inter-population spike correlations}

To understand how the cross-frequency coupling affects the underlying spiking dynamics, we simulated the spiking dynamics (Eq. \ref{eq:syn-homo-N-1} and Eq. \ref{eq:theta}) in the case of CFC and analyzed the spike correlations. Parameters are the same as  in Fig. \ref{fig:Delay-coupled-populations-of} (f,g). We obtain the spike times of the $i$ th neuron in $X$ population as $ t^{X_{(i)}}_1,\,t^{X_{(i)}}_2,\,t^{X_{(i)}}_3,...$ from the simulation. The spike trains are defined as $y_{i}^{X}(t)=\Sigma_{k}\delta(t-t^{X_{(i)}}_k)$. For the method of spike correlation analysis, we basically followed previous work \citep{Slow_dynamics}.

\begin{figure}[htp]

\includegraphics[width=1\linewidth]{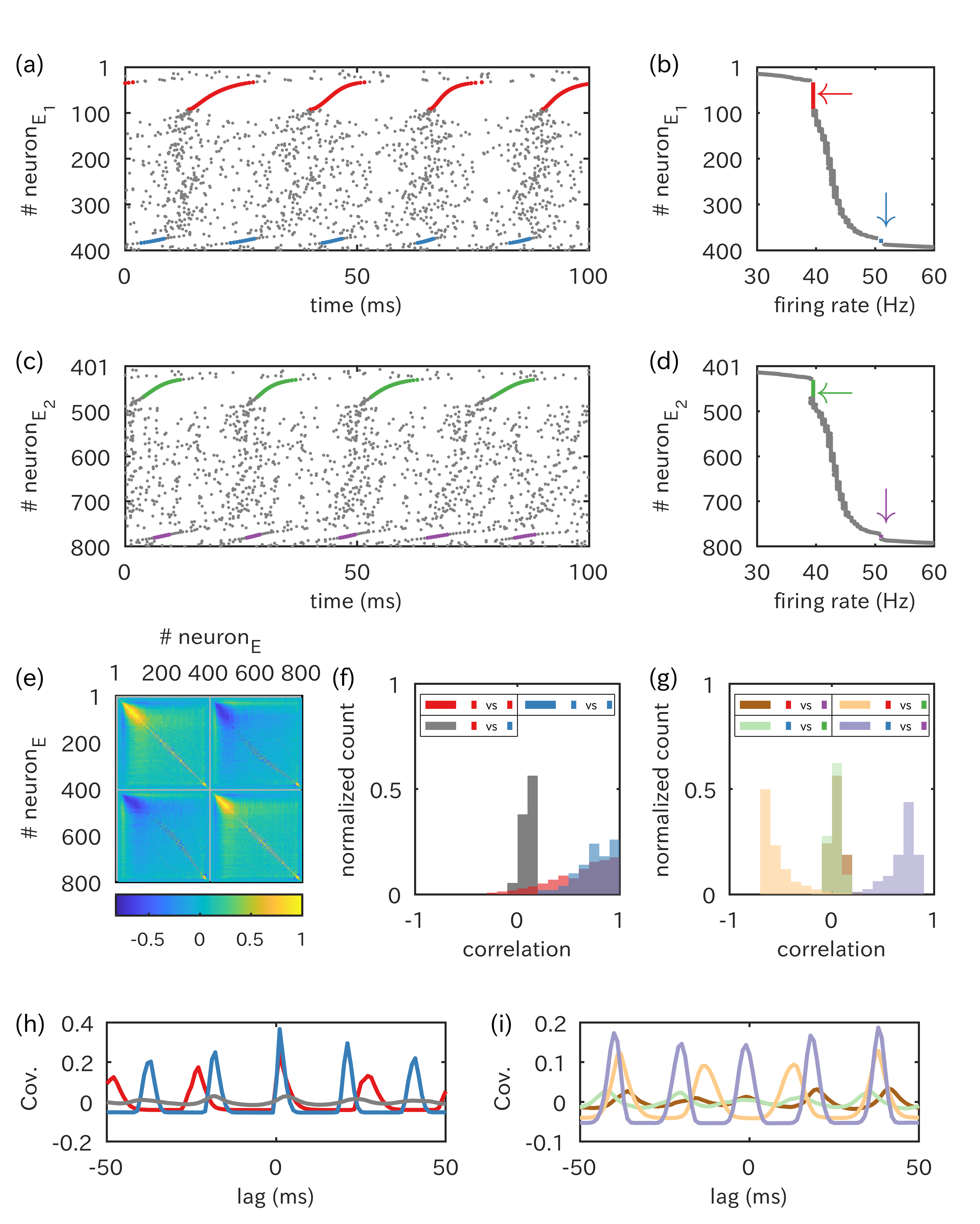}

\caption{CFC exibits long-range multi-cluster gamma-band cross-covariance (a) Rastergram of $E_{1}$. Two clusters are colored as red and blue. (b) Firing rate of each neuron. Note that the firing rate is the same within each cluster. (c) Rastergram of $E_{2}$. Two clusters are colored as green and purple. (d) Firing rate of each neuron. Similarly, the firing rate is the same within each clusters. (e) Correlation coefficients for each neuron pair. Time bin is 10(ms). (f) Histogram of correlation coefficients for pairs within red clusters (red), pairs within blue clusters (Blue) and pairs between red and blue clusters (Gray) in $E_{1}$. We can see firings within each clusters are highly correlated, while there is low correlation between the clusters. (g) Histogram of correlation coefficients between clusters belonging to different populations. Red-green pair and blue purple pairs are highly correlated, while red-purple and blue-green pairs exhibit low correlations. (h) Averaged cross-covariance function for corresponding pairs in (f). Time bin is 1(ms). Both red and blue clusters exhibit oscillatory firing. Note that the frequency is slightly different, which results in low cross-covariance between red and blue clusters. (i) Averaged cross-covariance function for corresponding pairs in (g). Time bin is 1(ms). Red-green pairs and blue-purple pairs, which are located in distant neuronal modules, are oscillating with the same frequency and exhibit high correlation.\label{fig:clusters}}
\end{figure}

The rastergram of the $E_{1}$ population is shown in Fig. \ref{fig:clusters}(a). Neurons are sorted in ascending order of $I_{i}^{E_{1}}$. We can see two lines running periodically in the rastergram (colored as red and blue). The firing rate of each neuron in $E_{1}$ is shown in Fig. \ref{fig:clusters}(b). Note that the firing rate is the same within red and blue clusters, respectively (Red cluster: $34 \leq i \leq 93$, Blue cluster: $375 \leq i \leq 384$). In Fig. \ref{fig:clusters}(c) and \ref{fig:clusters}(d), the same analysis is performed for $E_{2}$ . Similarly, we found two lines running periodically (colored as green and purple) and the firing rates are the same within each cluster (Green cluster: $431 \leq i \leq 470$, Purple cluster: $774 \leq i \leq 781$).

To quantify the relationship between the spike trains, we computed $\rho_{ij}^{XY}$, which is correlation coefficient between the $i^{th}$ neuron in $X$ and the $j^{th}$ neuron in $Y$ as

\begin{equation}
\rho_{ij}^{XY}=\frac{\mathrm{Cov}\left(N_{i}^{X}\left(t,t+\Delta t\right),N_{j}^{Y}\left(t,t+\Delta t\right)\right)}{\sqrt{\mathrm{Var}\left(N_{i}^{X}\left(t,t+\Delta t\right)\right)\mathrm{Var}\left(N_{j}^{Y}\left(t,t+\Delta t\right)\right)}},
\end{equation}

where $N_{i}^{X}(t,t+\Delta t)$ is the number of spikes emitted by the $i^{th}$ neuron in $X$ population during a time bin $\Delta t$, which is given by

\begin{equation}
N_{i}^{X}\left(t,t+\Delta t\right)=\int_{t}^{t+\Delta t}y_{i}^{X}(t')dt'.
\end{equation}

In our study, the time bin $\Delta t$ was set as $10$ (ms). Fig. \ref{fig:clusters}(e) displays the pairwise correlations for all of the pairs within and between $E_1$ and $E_2$. We can visually confirm high correlation within the red and green clusters, as well as negative correlation between the red and green clusters.

We further plotted histograms of correlation coefficients as shown in Figs. \ref{fig:clusters}(f) and \ref{fig:clusters}(g). In Fig. \ref{fig:clusters}(f), we plotted histograms of correlations within red clusters (red), within blue clusters (blue), and between red and blue clusters (gray) in $E_{1}$ population. We can see the firing within red and blue clusters is highly correlated, while low correlation is shown between the red and blue clusters, which depicts the local decorrelation of the clusters. We also plotted the histogram of the correlation coefficients between the clusters located in different populations ($E_{1}$ and $E_{2}$) in Fig. \ref{fig:clusters}(g). Red-green pairs and blue-purple pairs are highly correlated, while red-purple and blue-green pairs exhibit low correlations. The results indicate that highly correlated spiking activity, which is known as a typical index of spike transmission, emerges across regions in individual or sub-cluster level via the time-delay in communication.

Although some cluster pairs are shown to be decorrelated from the histogram, 
one might overlook temporally-localized correlated activity between the clusters. To further investigate the temporal structure of the correlation, we derived $C_{ij}^{XY}(\tau)$ which is a cross-covariance function between the $i^{th}$ neuron in $X$ and the $j^{th}$ neuron in $Y$ as

\begin{align}
C_{ij}^{XY}(\tau)= & \sum_{n}Y_{i}^{X}\left(t_{n}\right)Y_{j}^{Y}\left(t_{n}-\tau\right)\nonumber \\
 & -\frac{1}{\Delta T}N_{i}^{X}(t,t+\Delta T)\frac{1}{\Delta T}N_{j}^{Y}(t,t+\Delta T),
\end{align}

where $t_{n}=n\delta t$ and $Y_{i}^{X}(t_{n})=\int_{t_{n}}^{t_{n}+\delta t}y_{i}^{X}(t')dt'$ is the time binned spike train. The time bin $\delta t$ was set as $1$(ms) and $\Delta T$ was $2$(sec). Then, the averaged cross-covariance between the colored clusters is derived as the averaged $C_{ij}^{XY}(T)$ over the corresponding neuron pairs.

In Fig. \ref{fig:clusters}(h), we can see the cross-covariance between red and blue clusters is relatively small (gray), in relation to the pairs within clusters (red and blue), which shows the two clusters are decorrelated although they belong to the same excitatory population ($E_{1}$). We note that the both red and blue clusters show oscillatory firing in gamma band with slightly different frequencies (red: around 40Hz, blue: around 50Hz). We also evaluated cross-covariance functions for the inter-regional pairs as shown in Fig. \ref{fig:clusters}(i). Red-green pairs are oscillating with the same frequency (around 50 Hz) and therefore highly correlated. Similarly, blue-purple pairs are oscillating with the same frequency (around 40 Hz).
We note that blue-purple pairs are positively correlated while red-green pairs are negatively correlated in Fig. \ref{fig:clusters}(g). This is the consequence of the phase lag between them as seen in Fig. \ref{fig:clusters}(i).

These results indicate that the long range delayed coupling entrains the neurons in different regions to fire in the same frequencies in the gamma-band and generate cross-covariance across the distant brain regions. Also, the gamma-band entrainment can occur with several gamma-band pairs at the same time, which results in the beat between the gamma-bands and resulting CFC. 

\subsection{Suitable temporal window for population rate code}

It is experimentally reported that neurons tend to fire in specific phases of ongoing 
background oscillations\citep{Distinct_pop}, and this relation between phase of oscillation and spike timing is known to encode information\citep{theta_prec}. To evaluate any relationship between the phase of the background oscillation and spike timing, we introduced the mean membrane potential to evaluate the phase of the background oscillation and investigated the spike counts in relation to these phases in $E_{1}$ population.
There are two possible measures to evaluate the mean excitability of a neuronal population: population spike count and mean membrane voltage. In this study, we used the mean membrane potential because it is a smoother function of time than population spike counts. (It can be defined as a continuous function, while the population spike count is discrete.) In order to avoid the effect of physiologically implausible blow-up of membrane potential, we obtained the membrane potential of each neuron as: $V_{i}^{X}=\left(V_{R}+V_{T}\right)/2+\left(\sin\theta_{i}^{X}/(1+\cos\theta_{i}^{X}+\epsilon)\right)\left(V_{T}-V_{R}\right)/2$. A small constant $\epsilon=2.0\cdot10^{-4}$ is introduced in the denominator in order to avoid the divergence to infinity at $\theta=\pm\pi$~\citep{kotani_ing,gap_destroy}. Then, from the time course of individual neuron data given by Eq.\ref{eq:syn-homo-N-1} and Eq.\ref{eq:theta}, the mean membrane potential $ \langle V \rangle (t)$ is evaluated as

\begin{align}
 & \langle	V　\rangle	(t) =\frac{1}{N_{E_{1}}} \sum_{i=1}^{N_{E_{1}}}V_{i}^{E_{1}}.\label{eq:V_pop}
\end{align}

\begin{figure}[htp]

\includegraphics[width=1\linewidth]{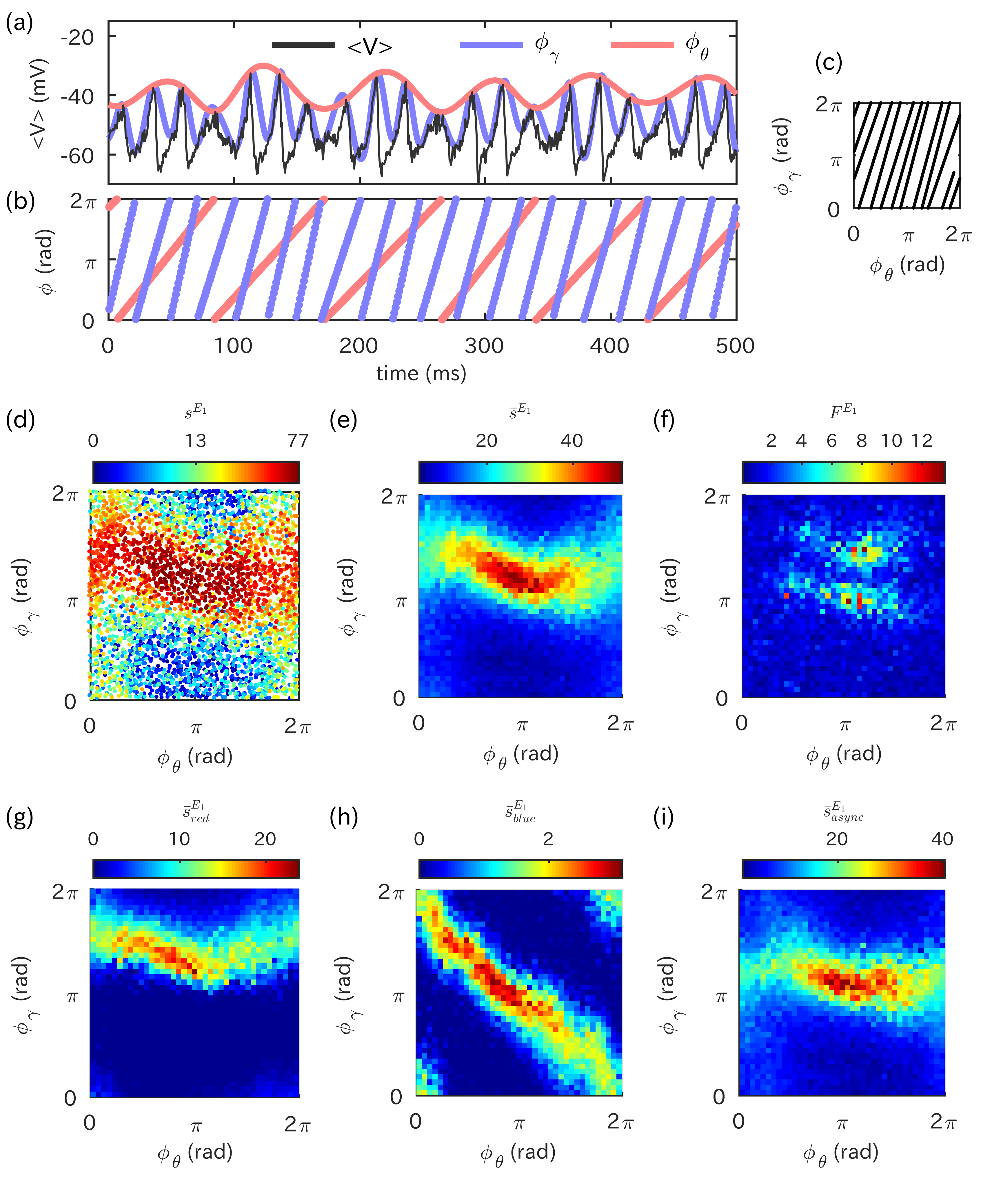}

\caption{$\theta$ -$\gamma$ oscillations and spiking dynamics. (a) Time series of $ \langle V \rangle (t)$, the mean of the membrane potential of neurons in $E_{1}$(Black) and the envelopes of $ \langle V \rangle (t)$ for theta oscillations (red) and gamma oscillations (blue). (b) Introducing $\phi_{\theta}$ and $\phi_{\gamma}$. (c) Time evolution of $\phi_{\theta}$ and $\phi_{\gamma}$. (d) Population spike count plotted over $\phi_{\theta}$ - $\phi_{\gamma}$ space. (e) Averaged spike counts for each bin in $\phi_{\theta}$-$\phi_{\gamma}$ space. (f) Fano factors of spike count. (g-i) Averaged spike counts of the sub-clusters found in Fig. \ref{fig:clusters}. (g) The red cluster (h) The blue cluster. (i) The neurons between the red and blue clusters. \label{fig:theta-gamma}}
\end{figure}

Based on $ \langle V \rangle (t)$ , we derived the envelopes and defined macroscopic phase for the theta oscillation ($\phi_{\theta}$) and the gamma oscillation ($\phi_{\gamma}$). In Fig. \ref{fig:theta-gamma} (a), the time series of $ \langle V \rangle (t)$ and its envelopes are shown. Envelopes are captured as the amplitude of the analytic signal of $ \langle V \rangle (t)$, which was derived via Hilbert filters. To capture the theta and gamma phase, we applied two Hilbert filters with different filter length (250ms for $\theta$, 50ms for $\gamma$)~\citep{hilbelt_time_length}. The zero phase for $\phi_{\theta}(t)$ and $\phi_{\gamma}(t)$ is defined as  the negative peak of the envelope as shown in Fig. \ref{fig:theta-gamma} (b). Because the macroscopic dynamics is a torus, where the frequency components are rationally independent generically, then the trajectory covers the whole area of the $\phi_{\theta}$-$\phi_{\gamma}$- 2D space as shown in Fig \ref{fig:theta-gamma} (c).

We plotted whole population spike counts $s^{E_{1}}\left(\phi_{\theta}\left(t\right),\phi_{\gamma}\left(t\right)\right)=\sum_{i}^{N_{E_{1}}}N_{i}^{E_{1}}\left(t-\delta t/2,t+\delta t/2\right)$ on the 2D space as shown in Fig. \ref{fig:theta-gamma} (d). We can see the spike count is modulated by both $\phi_{\theta}$ and $\phi_{\gamma}$.

To further investigate how the spike count is modulated by the phases,
we introduced $M$-binned phases $\phi^{m}=2m\pi/M$ in $\phi_{\theta}$-$\phi_{\gamma}$- 2D space 
and derived averaged spike counts for each bin
$\bar{s}^{E_{1}}\left(\phi_{\theta}^{i},\phi_{\gamma}^{j}\right)=
\mathrm{E}
\left(s^{E_{1}}\left(\phi_{\theta}^{i},\phi_{\gamma}^{j}\right)\right)$
where $\phi^{i}\leq\phi_{\theta}^{i}<\phi^{i+1}$ and $\phi^{j}\leq\phi_{\gamma}^{j}<\phi^{j+1}$ as shown in Fig. \ref{fig:theta-gamma} (e).
Here, the number of bins is $M=40$.
In Fig. \ref{fig:theta-gamma} (e), there is a region of high $\bar{s}^{E_{1}}$.
In addition, to evaluate the variability of the spike count,
we derived Fano factors for the spike count $F^{E_{1}}\left(\phi_{\theta}^{i},\phi_{\gamma}^{j}\right)=\mathrm{Var}\left(s^{E_{1}}\left(\phi_{\theta}^{i},\phi_{\gamma}^{j}\right)\right)/\bar{s}^{E_{1}}\left(\phi_{\theta}^{i},\phi_{\gamma}^{j}\right)$ \citep{Slow_dynamics}, for each bin as shown in Fig. \ref{fig:theta-gamma} (f).
From these calculations, we can see that the region with high $\bar{s}^{E_{1}}$ in Fig.\ref{fig:theta-gamma}(e) is interposed by the two regions with high $F^{E_{1}}$ as shown in Fig. \ref{fig:theta-gamma}(f). Between the region with high $F^{E_{1}}$, we can see a narrow area in which $\bar{s}^{E_{1}}$ is high and $F^{E_{1}}$ is low. We note that for a population rate code, we would like a high spike count average and a low Fano factor.

We also investigated how the temporal firing activity of the sub-clusters, which was found in Fig. \ref{fig:clusters} (a), is modulated by the macroscopic phases. The averaged spike count for the red cluster (
$\bar{s}^{E_{1}}_{red}$ : $34 \leq i \leq 93$), the blue cluster ($\bar{s}^{E_{1}}_{blue}$ : $375 \leq i \leq 384$) and the asynchronous neurons between the red and blue clusters ($\bar{s}^{E_{1}}_{async}$ : $94 \leq i \leq 374$) are shown in Figs. \ref{fig:theta-gamma} (g) ,(h) and (i). We can see these clusters exhibit several distinct types of phase-specific firing modulated by both $\phi_{\theta}$ and $\phi_{\gamma}$

\section{Discussion}

We employed numerical and theoretical analyses for inter-regionally coupled neuronal populations in a multi-scale point of view. We found the emergence of nontrivial CFC, induced by time-delay, with correlated spike trains between regions. We introduced quenched variability to individual neurons, instead of individual noise, in order to analyze the effect of time-delayed interactions thus avoiding the non-Markov state where noise and delay coexist.

Another key technique is that we consider conductance based synapses and voltage dependent neuronal dynamics. By the Lorentzian ansatz \citep{montbrio_OA_QIF} (it is associated with the Ott-Antonsen ansatz \citep{Ott_Antonsen_OA}), the dynamics of neuronal populations reduce to a set of macroscopic DDEs. These enable us to unveil the significant impact of the delay on the rhythmogenesis. As the derived equations are valid regardless of the frequencies, we successfully investigate theta-gamma interactions, which is not possible by conventional firing-rate models~\citep{Bard_mathematical_foundataion,temporal_deccoration}. In addition, the mean (Eq. \ref{eq:V_pop}) and individual voltage (Eq. \ref{eq:V_N}) of neurons are appropriately evaluated under biologically plausible values for the conductances and reversal potentials of GABA and AMPA~\citep{brunel_what_determines}.

By the analyses of the macroscopic equations, we found that time-delay destabilizes the asynchronous firing state via a Hopf bifurcation (HB). Further increase of the time-delay leads to a destabilization of the oscillation (that emerged from the HB) via a torus bifurcation which produces the CFC. Fig. \ref{fig:Bifurcation-diagrams} (b) shows that the torus bifurcation emerges at the intersection of two curves of HBs. \cite{buono} performed a partial analysis of the case in which there is a double HB in a delay model in the non-resonant case (when the frequencies of the HBs were not rationally related); in a broad range of parameters, it is possible to find a stable torus that emerges. (See also \cite{ErmentroutCowan80} for a neuronal example.)
This appears to be the case in our system. Under such bifurcations, diverse interactions between the macroscopic oscillation and individual neurons can be realized. By analyses of the microscopic spike trains under CFC, we found that the inter-population correlation can be much larger than that of the intra-population. We also found that there is a firing preference of sub-populations as a function of the phase of the slow oscillation.

In previous studies, the gamma oscillation were found to emerge from local inhibitions~\citep{buz-wang,kotani_ing}. Our findings here show that a gamma oscillation can emerge from inter-regional excitatory time-delayed interactions (e.g. Fig. \ref{fig:Bifurcation-diagrams} (a)). This delay-induced mechanism is different from the conventional mechanisms (ING and PING), thus implies a novel generation mechanism for the gamma oscillation.


Pairwise spike correlation has been widely analyzed in physiological experiments. It is reported to increase or decrease with respect to task demands\citep{state_dependent_corr}. Although the relation of correlations to brain functions has been experimentally suggested, it is still hard to interpret them and understand the origin of such correlations. As for the dynamical mechanism, Litwin-Kumar et al. proposed that a slight heterogeneity of neural interaction increased pairwise correlation in the same cluster~\citep{Slow_dynamics}. In our model, there are two clusters in each population that show the same firing rate. High correlations in absolute value emerge between neurons within different populations, that are located in distant regions, while low correlation occur between populations that have different firing rates. CFC induced by time delay can thus serve multi-band information transfer by using sub-cluster correlations and may support neuronal multiplexing \citep{Akam_multiplex}.

Regarding the spike preferences of macroscopic phase, Georgiou et al. show that the spike timing of neurons in V4 and the frontal eye fields occurs at specific phases of LFP in V4 during visual attentions~\citep{LR_gamma_LPF_spike}. 
Sellers et al. also found that neurons in Prefrontal cortex fire at specific phases of not only local theta and gamma oscillation, but also phase of theta oscillation of anatomically connected distant region~\citep{LR_thetagamma_LFP_spike}.
These studies suggest that the spike preference of macroscopic phase could encode and transfer information. In Fig. \ref{fig:theta-gamma}, our model also exhibits spike preference of the macroscopic phase of theta and gamma oscillations.

Senior et al. evaluated firing timing of hippocampal CA1 pyramidal cells and found phase preferences in relation to theta and gamma oscillations~\citep{Distinct_pop}. They also found distinct two types neuron groups that have a different phase preference (Fig. 4 B in \citep{Distinct_pop}) and firing rate (Fig. 2 B in \citep{Distinct_pop}). 
Such clusters, which exhibit distinct phase preference and firing rates, were also found in our model during CFC. Moreover, we found these clusters exhibit long-range gamma-band cross-covariance across distant regions, suggesting the impact on long-range information transfer. We note the emergence of such sub-clusters in a population, which were characterized by different firing rates, are also reported even under collective chaos \citep{Pazo_QPS2chaos_delay,plateus}. Further studies are needed to figure out 
whether sub-clusters can also contribute to long-range communication in a synchronized manner in chaotic regions, where population level signals are hard to correlate \citep{temporal_deccoration}.


The theta-gamma neural code hypothesis, proposed by Lisman et al., assumes 
that the nested oscillation, where the phase of theta oscillation and the amplitude of gamma oscillation are coupled, encode multiple memory items in an ordered way\citep{lisman_tgcode}.
Although this hypothesis is supported by recent experimental studies~\citep{hsusser_tgcode}, it is less clear whether the dynamical properties of such nested oscillations have the ability to store information.  We showed, in Figs. \ref{fig:theta-gamma}(e) and \ref{fig:theta-gamma}(f), that nested oscillations accompany a  phase-specific time window with high-spike counts and low-Fano factor. Within such a temporal window, stimuli are likely to be robustly encoded by changes of spike counts.


In real brains, interactions between multiple regions would exhibit diverse time-delayed interactions~\citep{delay_deco}.  Further investigation about delayed interactions and information transmission will be important in order to understand how spike dynamics is affected by the macroscopic phase of multiple brain regions~\citep{Large_scale_cell_assembly}. 

\begin{acknowledgments}
This study was supported in part by Grant-in-Aid for JSPS Research Fellow Grant No. 16J04952 to A.A. K.K. was supported by JST PRESTO (JPMJPR14E2) and KAKENHI (18H04122) B.E. was supported by the US NSF DMS-1712922.
We would like to thank Brent Doiron for helpful discussions.
\end{acknowledgments}


\appendix

\section{Derivation of mean field equations\label{subsec:Derivation-of-mean}}

In this section, the details of the derivation of mean field equations are described. In the derivation, we mostly follow the previous work by Montbri\'{o} et al \citep{montbrio_OA_QIF}.

\subsection{Spiking description to reduce by Ott-Antonsen ansatz}

The dynamics of the membrane potential $V_{i}^{X}(t)$ , $i$-th neuron in population $X$, is

\begin{align}
C\frac{dV_{i}^{X}}{dt}= & g_{LX}\frac{(V_{i}^{X}(t)-V_{R})(V_{i}^{X}(t)-V_{T})}{V_{T}-V_{R}}\nonumber \\
 & -\sum_{Y}g_{X}^{Y}(t)(V_{i}^{X}(t)-V_{syn}^{Y})+I_{i}^{X},\label{eq:V_N-1}
\end{align}

which can be written as

\begin{align}
\frac{dV_{i}^{X}}{dt}= & a_{X}\left(V_{i}^{X}(t)\right)^{2}+b_{X}(t)V_{i}^{X}(t)+c_{X}(t)+I_{i}^{X},\label{eq:V_N-1-1-1}
\end{align}

where $a_{X}=g_{LX}/(V_{T}-V_{R})$, $b_{X}(t)=-g_{LX}(V_{T}+V_{R})/(V_{T}-V_{R})-\sum_{Y}g_{X}^{Y}(t)$ and $c_{X}(t)=-g_{LX}V_{T}V_{R}/(V_{T}-V_{R})+\sum_{Y}g_{X}^{Y}(t)V_{syn}^{Y}$.

Also, the dynamics of $g_{X}^{Y}(t)$ is written as 
\begin{equation}
\frac{dg_{X}^{Y}}{dt}=-\frac{1}{\tau_{d}^{Y}}g_{X}^{Y}(t)+\bar{g}_{X}^{Y}\cdot p_{X}^{Y}\cdot\sum_{k=1}^{N_{Y}}\delta(t-(t^{(k^{Y})}+\tau_{X}^{Y})).\label{eq:syn-homo-N}
\end{equation}

Here, the N-body description of the whole system is 1012-dimensional system with 1000 of $V_{i}^{X}$ and 12 of $g_{X}^{Y}$.

\subsection{Introducing probability density function $\rho_{X}\left(V_{X}|I_{X},t\right)$ }

Taking the continuum limit as $N_{X}\rightarrow\infty$, we introduce the probability density function $\rho_{X}(V_{X}|I_{X},t)$ for the population $X$, where $\int_{\nu}^{\nu+\Delta\nu}\rho_{X}\left(V_{X}|I_{X},t\right)dV_{X}$ describes the probability of neurons in the population whose membrane potentials $V_{X}$ are between $\nu$ and $\nu+\Delta\nu$ and current is $I_{X}$ at time $t$. Because the number of neuron is conserved, $\rho_{X}(V_{X}|I_{X},t)$ follows the continuity equation: 
\begin{align}
\frac{\partial}{\partial t}\rho_{X}(V_{X}|I_{X},t)\nonumber \\
=-\frac{\partial}{\partial V_{X}} & \left[\left(a_{X}V_{X}^{2}+b_{X}(t)V_{X}+c_{X}(t)+I_{X}\right)\right.\nonumber \\
 & \left.\rho_{X}(V_{X}|I_{X},t)\right]
\end{align}

\subsection{Adopting the ``Ott-Antonsen ansatz''}

Here, we start to derive the mean field dynamics of the system by applying so-called ``Ott-Antonsen ansatz (OAA)''. In the previous study\citep{montbrio_OA_QIF}, the ansatz is extended to Quadratic integrate-and-fire neurons as:

\begin{equation}
\rho_{X}\left(V_{X}|I_{X},t\right)=\frac{f(I_{X})}{\pi}\frac{x_{X}(I_{X},t)}{\left[V_{X}-y_{X}(I_{X},t)\right]^{2}+x_{X}(I_{X},t)^{2}},
\end{equation}

which is a Lorentzian distribution with dynamical variables 
$x_{X}(I_{X},t)$ and  $y_{X}(I_{X},t)$.
Here, $x_{X}(I_{X},t)$ and $y_{X}(I_{X},t)$ represent the low dimensional behavior of the probability density function $\rho_{X}$. Adopting the ansatz, we obtain the low dimensional behavior as 
\begin{equation}
\frac{d}{dt}x_{X}(I_{X},t)=2a_{X}x_{X}(I_{X},t)y(I_{X},t)+b_{X}(t)x_{X}(I_{X},t),
\end{equation}
\begin{align}
\frac{d}{dt}y_{X}(I_{X},t)= & -a_{X}\left(x_{X}(I_{X},t)\right)^{2}+a_{X}\left(y_{X}\left(I_{X},t\right)\right)^{2}\nonumber \\
 & +b_{X}(t)y_{X}\left(I_{X},t\right)+c_{X}(t)+I_{X}.
\end{align}

Introducing the complex variable $w_{X}(I_{X},t)=x_{X}(I_{X},t)+iy_{X}(I_{X},t)$, the two coupled equations can be written in complex form as

\begin{equation}
\frac{d}{dt}w_{X}(I_{X},t)=ia_{X}w_{X}(I_{X},t)^{2}+b_{X}(t)w_{X}(I_{X},t)+c_{X}(t)+I_{X}.\label{eq:dunamics-of-alpha}
\end{equation}

\subsection{Description of mean field dynamics: $r_{X}(t)$ and $v_{X}(t)$}

We introduce two macroscopic observables: the firing rate $r_{X}(t)$ and the mean voltage $v_{X}(t)$. The firing rate of the population $r_{X}(t)$ is obtained by summing up the flux for all $I_{X}$ at $V_{X}=V_{peak}$ , where $V_{peak}$ is the firing threshold. Taking the firing threshold $V_{peak}\rightarrow\infty$, $r_{X}(t)$ can be defined as

\begin{align}
r_{X}(t)= & \int_{-\infty}^{\infty}\lim_{V_{X}\rightarrow\infty}(a_{X}V_{X}^{2}+b_{X}(t)V_{X}+c_{X}(t)+I_{X})\nonumber \\
 & \rho_{X}(V_{X},I_{X},t)dI_{X}.
\end{align}

Substituting the ansatz, we get

\begin{align}
r_{X}(t) & =\frac{a_{X}}{\pi}\int_{-\infty}^{\infty}x_{X}(I_{X},t)f_{X}(I_{X})dI_{X}.\label{eq:r_define-1}
\end{align}

Following that $f_{X}(I_{X})$ is now given as the Lorentzian function, this improper integration can be evaluated using an analytic continuation and the residue theorem, then we get

\begin{align}
r_{X}(t) & =\frac{a_{X}}{\pi}x_{X}(\bar{I}_{X}-i\Delta_{I_{X}},t).\label{eq:r_define-1-1}
\end{align}

Next, the mean voltage of the population $v(t)$ can be defined by integrating the $V_{X}$-weighted $\rho_{X}(V,I,t)$ for all the $V_{X}$ and $I_{X}$ values, so that

\begin{align}
v_{X}(t) & =\int_{-\infty}^{\infty}\mathrm{p.v.}\int_{-\infty}^{\infty}\rho_{X}(V_{X},I_{X},t)V_{X}dV_{X}dI_{X}.\label{eq:v_define}
\end{align}

Note that we resort to the Cauchy principal value $\mathrm{p.v.}\int_{-\infty}^{\infty}h(x)dx=\lim_{R\rightarrow\infty}\int_{-R}^{R}h(x)dx$ in order to avoid indeterminacy of the improper integral.
Substituting the ansatz, we get

\begin{align}
v_{X}(t) & =\int_{-\infty}^{\infty}y_{X}(I_{X},t)f_{X}(I_{X})dI_{X}.\label{eq:r_define-1-2}
\end{align}

As the same as $r_{x}(t)$ case, following that $f_{X}(I)$ is now given as the Lorentzian function, this improper integration can be evaluated using the residue theorem to get

\begin{align}
v_{X}(t) & =y_{X}(\bar{I}_{X}-i\Delta_{I_{X}},t).\label{eq:r_define-1-1-1}
\end{align}

Finally, substituting Eq. \ref{eq:r_define-1-1} and Eq. \ref{eq:r_define-1-1-1} into Eq.\ref{eq:dunamics-of-alpha}, the population dynamics is obtained as

\begin{align}
\frac{d}{dt}r_{X}(t) & =2a_{X}r_{X}(t)v_{X}(t)+b_{X}(t)r_{X}(t)+\frac{a_{X}}{\pi}\Delta_{I_{X}},\label{eq:r-dynamics-1}
\end{align}

\begin{align}
\frac{d}{dt}v_{X}(t) & =-\frac{\pi^{2}}{a_{X}}r_{X}(t)^{2}+a_{X}v_{X}(t)^{2}+b_{X}(t)v_{X}(t)+c_{X}(t)+\bar{I}_{X}.\label{eq:v-dynamics-1}
\end{align}

\subsection{Description of mean field synaptic dynamics: $g_{X}^{Y}(t)$}

Since we have the dynamics of the population firing rate $r_{X}(t)$, Eq. \ref{eq:syn-homo-N} can be written using $r_{X}(t)$ instead of using delta function as 
\begin{equation}
\frac{dg_{X}^{Y}}{dt}=-\frac{1}{\tau_{d}^{Y}}g_{X}^{Y}(t)+\bar{g}_{X}^{Y}\cdot p_{X}^{Y}\cdot N_{Y}\cdot r_{Y}(t-\tau_{X}^{Y}).\label{eq:syn-homo-N-1-1}
\end{equation}


\begin{thebibliography}{99}

\bibitem{buzsaki_book}Buzs\'{a}ki, Gy\"{o}rgy. Rhythms of the Brain. Oxford University Press, 2006.

\bibitem{wang_rev}Wang, Xiao-Jing. \textquotedbl{} Neurophysiological and computational principles of cortical rhythms in cognition.\textquotedbl{} Physiological reviews 90.3 (2010): 1195-1268.
\bibitem{buz-wang} Buz\'{a}ki,Gy\"{o}rgy and Xiao-jing Wang, Mechanisms of gamma oscillations. Annual review of neuroscience 35 (2012): 203-225.

\bibitem{Jensen_CFC_rev}Jensen, Ole, and Laura L. Colgin. \textquotedbl{} Cross-frequency coupling between neuronal oscillations.\textquotedbl{} Trends in cognitive sciences 11.7 (2007): 267-269.

\bibitem{canolty_CFC_rev}Canolty, Ryan T., and Robert T. Knight. \textquotedbl{} The functional role of cross-frequency coupling.\textquotedbl{} Trends in cognitive sciences 14.11 (2010): 506-515.

\bibitem{Juhan_CFC_rev}Aru, Juhan, et al. "Untangling cross-frequency coupling in neuroscience." Current opinion in neurobiology 31 (2015): 51-61.

\bibitem{mizureki_CFC_tgcouple}Belluscio, Mariano A., et al. \textquotedbl{} Cross-frequency phase--phase coupling between theta and gamma oscillations in the hippocampus.\textquotedbl{} Journal of Neuroscience 32.2 (2012): 423-435.

\bibitem{CFC_Gutkin_rev} Hyafil, Alexandre, et al. "Neural cross-frequency coupling: connecting architectures, mechanisms, and functions." Trends in neurosciences 38.11 (2015): 725-740.


\bibitem{CFC_sase} Sase, Takumi, et al. "Bifurcation analysis on phase-amplitude cross-frequency coupling in neural networks with dynamic synapses." Frontiers in computational neuroscience 11 (2017): 18.

\bibitem{CFC_Hyafil_speech} Hyafil, Alexandre, et al. "Speech encoding by coupled cortical theta and gamma oscillations." Elife 4 (2015): e06213.


\bibitem{CFC_plone_1} Onslow, Angela CE, Matthew W. Jones, and Rafal Bogacz. "A canonical circuit for generating phase-amplitude coupling." PLoS One 9.8 (2014): e102591.

\bibitem{CFC_plone_2} Chehelcheraghi, Mojtaba, et al. "A neural mass model of cross frequency coupling." PloS one 12.4 (2017): e0173776.

\bibitem{brainweb}Varela, Francisco, et al. \textquotedbl{} The brainweb: phase synchronization and large-scale integration.\textquotedbl{} Nature reviews neuroscience 2.4 (2001): 229.

\bibitem{IR_CFC_1}Fontolan, Lorenzo, et al. \textquotedbl{} The contribution of frequency-specific activity to hierarchical information processing in the human auditory cortex.\textquotedbl{} Nature communications 5 (2014): 4694.

\bibitem{IR_CFC_2}Doesburg, Sam M., et al. \textquotedbl{} Theta modulation of inter-regional gamma synchronization during auditory attention control.\textquotedbl{} Brain research 1431 (2012): 77-85.

\bibitem{IR_CFC_3}Dynamic cross-frequency couplings of local field potential oscillations in rat striatum and hippocampus during performance of a T-maze task Adriano B. L. Tort, Mark A. Kramer, Catherine Thorn, Daniel J. Gibson, Yasuo Kubota, Ann M. Graybiel, and Nancy J. Kopell

\bibitem{IR_CFC_4}Canolty, Ryan T., et al. "High gamma power is phase-locked to theta oscillations in human neocortex." science 313.5793 (2006): 1626-1628.

\bibitem{Breakspear}Breakspear, Michael. \textquotedbl{} Dynamic models of large-scale brain activity.\textquotedbl{} Nature neuroscience 20.3 (2017): 340.

\bibitem{WC_wcmodel}Wilson, Hugh R., and Jack D. Cowan. \textquotedbl{} Excitatory and inhibitory interactions in localized populations of model neurons.\textquotedbl{} Biophysical journal 12.1 (1972): 1-24.

\bibitem{Bard_mathematical_foundataion}Ermentrout, G. Bard, and David H. Terman. Mathematical foundations of neuroscience. Vol. 35. Springer Science \& Business Media, 2010.

\bibitem{Devalle_spike_sychrony}Devalle, Federico, Alex Roxin, and Ernest Montbri\'{o}. \textquotedbl{} Firing rate equations require a spike synchrony mechanism to correctly describe fast oscillations in inhibitory networks.\textquotedbl{} PLoS computational biology 13.12 (2017): e1005881.


\bibitem{temporal_deccoration}Battaglia, Demian, Nicolas Brunel, and David Hansel. "Temporal decorrelation of collective oscillations in neural networks with local inhibition and long-range excitation." Physical review letters 99.23 (2007): 238106.


\bibitem{Ott_Antonsen_OA}Ott, Edward, and Thomas M. Antonsen. \textquotedbl{} Low dimensional behavior of large systems of globally coupled oscillators.\textquotedbl{} Chaos: An Interdisciplinary Journal of Nonlinear Science 18.3 (2008): 037113.

\bibitem{montbrio_OA_QIF}Montbri\'{o}, Ernest, Diego Paz\'{o}, and Alex Roxin. \textquotedbl{} Macroscopic description for networks of spiking neurons.\textquotedbl{} Physical Review X 5.2 (2015): 021028.


\bibitem{Luke_complete_pulse}Luke, Tanushree B., Ernest Barreto, and Paul So. \textquotedbl{} Complete classification of the macroscopic behavior of a heterogeneous network of theta neurons.\textquotedbl{} Neural computation 25.12 (2013): 3207-3234.

\bibitem{Laing_Gap_junction}Laing, Carlo R. \textquotedbl{} Exact neural fields incorporating gap junctions.\textquotedbl{} SIAM Journal on Applied Dynamical Systems 14.4 (2015): 1899-1929.

\bibitem{Pazo_QPS2chaos_delay}Paz\'{o}, Diego, and Ernest Montbri\'{o}. \textquotedbl{} From quasiperiodic partial synchronization to collective chaos in populations of inhibitory neurons with delay.\textquotedbl{} Physical review letters 116.23 (2016): 238101.



\bibitem{Devalle_a_large}Devalle, Federico, Ernest Montbri\'{o}, and Diego Paz\'{o}. "Dynamics of a large system of spiking neurons with synaptic delay." Physical Review E 98.4 (2018): 042214.


\bibitem{Dumont_OA_PRC}Dumont, Gr\'{e}gory, G. Bard Ermentrout, and Boris Gutkin. \textquotedbl{} Macroscopic phase-resetting curves for spiking neural networks.\textquotedbl{} Physical Review E 96.4 (2017): 042311.

\bibitem{Paulso_time_varying}So, Paul, Tanushree B. Luke, and Ernest Barreto. \textquotedbl{} Networks of theta neurons with time-varying excitability: Macroscopic chaos, multistability, and final-state uncertainty.\textquotedbl{} Physica D: Nonlinear Phenomena 267 (2014): 16-26.

\bibitem{brunel_what_determines}Brunel, Nicolas, and Xiao-Jing Wang. \textquotedbl{} What determines the frequency of fast network oscillations with irregular neural discharges? I. Synaptic dynamics and excitation-inhibition balance.\textquotedbl{} Journal of neurophysiology 90.1 (2003): 415-430.

\bibitem{reversal_V_affects_gamma}Stiefel, Klaus M., et al. \textquotedbl{} Phase dependent sign changes of GABAergic synaptic input explored in-silicio and in-vitro.\textquotedbl{} Journal of computational neuroscience 19.1 (2005): 71-85.

\bibitem{revV_affect_2}Vida, Imre, Marlene Bartos, and Peter Jonas. \textquotedbl{} Shunting inhibition improves robustness of gamma oscillations in hippocampal interneuron networks by homogenizing firing rates.\textquotedbl{} Neuron 49.1 (2006): 107-117.

\bibitem{kotani_ing}Kotani, Kiyoshi, et al. \textquotedbl{} Population dynamics of the modified theta model: macroscopic phase reduction and bifurcation analysis link microscopic neuronal interactions to macroscopic gamma oscillation.\textquotedbl{} Journal of The Royal Society Interface 11.95 (2014): 20140058.

\bibitem{akao_pre}Akao, Akihiko, et al. \textquotedbl{} Relationship between the mechanisms of gamma rhythm generation and the magnitude of the macroscopic phase response function in a population of excitatory and inhibitory modified quadratic integrate-and-fire neurons.\textquotedbl{} Physical Review E 97.1 (2018): 012209.

\bibitem{innocenti_delay_length}Caminiti, Roberto, et al. \textquotedbl{} Evolution amplified processing with temporally dispersed slow neuronal connectivity in primates.\textquotedbl{} Proceedings of the National Academy of Sciences 106.46 (2009): 19551-19556.

\bibitem{aon_exp}Bloomfield, S. A., J. E. Hamos, and S. M. Sherman. \textquotedbl{} Passive cable properties and morphological correlates of neurones in the lateral geniculate nucleus of the cat.\textquotedbl{} The Journal of physiology 383.1 (1987): 653-692.

\bibitem{aon}D. A. McCormick and J. R. Huguenard, J. Neurophysiol. 68, 1384 (1992).

\bibitem{g_peak_1}M. Bartos, I. Vida, M. Frotscher, J. R. P. Geiger, and P. Jonas, J. Neurosci. 21, 2687 (2001).

\bibitem{g_peak_2}A. Gupta, Y. Wang, and H. Markram,

\bibitem{ddebif_1}K. Engelborghs, T. Luzyanina, and D. Roose, Numerical bifurcation analysis of delay differential equations using DDE-BIFTOOL, ACM Trans. Math. Softw. 28 (1), pp. 1-21, 2002.

\bibitem{ddebif_2}K. Engelborghs, T. Luzyanina, G. Samaey. DDE-BIFTOOL v. 2.00: a Matlab package for bifurcation analysis of delay differential equations. Technical Report TW-330, Department of Computer Science, K.U.Leuven, Leuven, Belgium, 2001.

\bibitem{ddebif_3}J. Sieber, K. Engelborghs, T. Luzyanina, G. Samaey, D. Roose: DDE-BIFTOOL Manual - Bifurcation analysis of delay differential equations. arxiv.org/abs/1406.7144.

\bibitem{ddebif_4}Sebastiaan Janssens: On a Normalization Technique for Codimension Two Bifurcations of Equilibria of Delay Differential Equations. Master Thesis, Utrecht University (NL), supervised by Yu.A. Kuznetsov and O. Diekmann, dspace.library.uu.nl/handle/1874/312252, 2010.

\bibitem{ddebif_5}Bram Wage: Normal form computations for Delay Differential Equations in DDE-BIFTOOL. Master Thesis, Utrecht University (NL), supervised by Y.A. Kuznetsov, dspace.library.uu.nl/handle/1874/296912, 2014.

\bibitem{ddebif_6}M. M. Bosschaert: Switching from codimension 2 bifurcations of equilibria in delay differential equations. Master Thesis, Utrecht University (NL), supervised by Y.A. Kuznetsov, dspace.library.uu.nl/handle/1874/334792, 2016.


\bibitem{LeeDelay} Lee, Wai Shing, Edward Ott, and Thomas M. Antonsen. "Large coupled oscillator systems with heterogeneous interaction delays." Physical review letters 103.4 (2009): 044101.


\bibitem{Lorenz}Candaten, Matteo, and Sergio Rinaldi. \textquotedbl{} Peak-to-peak dynamics: A critical survey.\textquotedbl{} International Journal of Bifurcation and Chaos 10.08 (2000): 1805-1819.

\bibitem{strogatz}Strogatz, Steven H. Nonlinear dynamics and chaos: with applications to physics, biology, chemistry, and engineering. CRC Press, 2018.

\bibitem{Slow_dynamics}Litwin-Kumar, Ashok, and Brent Doiron. \textquotedbl{} Slow dynamics and high variability in balanced cortical networks with clustered connections.\textquotedbl{} Nature neuroscience 15.11 (2012): 1498.

\bibitem{buono} Buono, Pietro-Luciano, and Jacques Belair. Restrictions and unfolding of double Hopf bifurcation in functional differential equations.  Journal of Differential Equations 189.1 (2003): 234-266.


\bibitem{double-hopf}Knobloch, E. \textquotedbl{} Normal form coefficients for the nonresonant double Hopf bifurcation.\textquotedbl{} Physics Letters A 116.8 (1986): 365-369.


\bibitem{Distinct_pop}Senior, Timothy J., et al. \textquotedbl Gamma oscillatory firing reveals distinct populations of pyramidal cells in the CA1 region of the hippocampus.\textquotedbl{} Journal of Neuroscience 28.9 (2008): 2274-2286.


\bibitem{theta_prec} O'Keefe, John, and Michael L. Recce. "Phase relationship between hippocampal place units and the EEG theta rhythm." Hippocampus 3.3 (1993): 317-330.

\bibitem{gap_destroy} Ermentrout, Bard. "Gap junctions destroy persistent states in excitatory networks." Physical Review E 74.3 (2006): 031918.

\bibitem{hilbelt_time_length} Although we arbitrary controlled the filter length as to capture the theta and gamma envelopes, we also confirmed these result are robust against the minor changes of the filter length.


\bibitem{ErmentroutCowan80} Ermentrout, G. B., and J. D. Cowan. Secondary bifurcation in neuronal nets. SIAM Journal on Applied Mathematics 39.2 (1980): 323-340.



\bibitem{state_dependent_corr}Doiron, Brent, et al. "The mechanics of state-dependent neural correlations." Nature neuroscience 19.3 (2016): 383.


\bibitem{Akam_multiplex}Akam, Thomas, and Dimitri M. Kullmann. "Oscillatory multiplexing of population codes for selective communication in the mammalian brain." Nature Reviews Neuroscience 15.2 (2014): 111.


\bibitem{LR_gamma_LPF_spike}Gregoriou, Georgia G., et al. \textquotedbl High-frequency, long-range coupling between prefrontal and visual cortex during attention.\textquotedbl{} science 324.5931 (2009): 1207-1210.

\bibitem{LR_thetagamma_LFP_spike}Sellers, Kristin K., et al. \textquotedbl Oscillatory dynamics in the frontoparietal attention network during sustained attention in the ferret.\textquotedbl{} Cell reports 16.11 (2016): 2864-2874.



\bibitem{plateus}Luccioli, Stefano, and Antonio Politi. "Irregular collective behavior of heterogeneous neural networks." Physical review letters 105.15 (2010): 158104.




\bibitem{lisman_tgcode}Lisman, John E., and Ole Jensen. \textquotedbl{} The theta-gamma neural code.\textquotedbl{} Neuron 77.6 (2013): 1002-1016.

\bibitem{hsusser_tgcode}Heusser, Andrew C., et al. \textquotedbl{} Episodic sequence memory is supported by a theta--gamma phase code.\textquotedbl{} Nature neuroscience 19.10 (2016): 1374.






\bibitem{delay_deco}Deco, G., Jirsa, V., McIntosh, A. R., Sporns, O., \& K{\"o}tter, R. (2009). Key role of coupling, delay, and noise in resting brain fluctuations. Proceedings of the National Academy of Sciences, pnas-0901831106. 























\bibitem{Large_scale_cell_assembly}Canolty, Ryan T., et al. \textquotedbl Oscillatory phase coupling coordinates anatomically dispersed functional cell assemblies.\textquotedbl{} Proceedings of the National Academy of Sciences (2010): 201008306.



\end{thebibliography}
\end{document}